\documentclass{hcig}

\usepackage{amsmath}
\usepackage{mathrsfs}
\usepackage{graphicx}
\usepackage{hyperref}
\usepackage[binary-units=true]{siunitx}
\usepackage{algorithm}
\usepackage{algpseudocode}
\usepackage{amsfonts,amssymb}
\usepackage{mathtools}
\usepackage{bm}
\usepackage{wrapfig}
\usepackage{xcolor}
\usepackage[nameinlink]{cleveref}
\usepackage{array,tabularx,booktabs}

\newcolumntype{C}{>{\centering\arraybackslash}X}
\newcolumntype{W}{>{\centering\arraybackslash\hsize=1.2\hsize}X}
\newcolumntype{S}{>{\centering\arraybackslash\hsize=.9\hsize}X}

\definecolor{hlcolor}{RGB}{220, 245, 220}

\def\lim{\mathop{\mathsf{lim}}} 

\def\log{\mathsf{log\,}}

\def\R{\mathbb{R}}
\def\E{\mathbb{E}}

\def\hbf{{\mathbf{h}}}

\def\xbf{{\mathbf{x}}}
\def\ybf{{\mathbf{y}}}
\def\zbf{{\mathbf{z}}}

\def\proposed{VOLT}

\title{\proposed: Volumetric Wide-Field Microscopy via 3D-Native Probabilistic Transport}

\author{Yetao He$^2$, Wenhan Guo$^1$, Deliang Wei$^1$, Evan Bell$^1$, Ji Yi$^{2,3}$, and Yu Sun$^{1,}$\textsuperscript{\Letter}}
\address{$^1$Department of Electrical and Computer Engineering, Johns Hopkins University \\ 
$^2$Department of Biomedical Engineering, Johns Hopkins University\\
$^3$Wilmer Eye Institute, Johns Hopkins Medicine\\
\smallskip
{\footnotesize Emails: \color{jhu}{\{yhe121, wguo28, dwei12, ebell34, ysun214\}@jh.edu, jiyi@jhu.edu}}\\
{\footnotesize \textsuperscript{\Letter}Corresponding author: \color{jhu}{ysun214@jh.edu}}}

\headertitle{\proposed}
\headerauthors{He et al.}

\begin{document}

\maketitle
\thispagestyle{firstpagestyle}

\begin{abstract}
Three-dimensional (3D) wide-field fluorescence microscopy is a widely used modality for volumetric imaging, but suffers from characteristic out-of-focus blur. 
Existing reconstruction methods either struggle to operate on high-dimensional volumes or fail to provide credibility characterization of the reconstruction.
In this work, we introduce Volumetric Transport (VOLT), a 3D-native probabilistic framework for wide-field fluorescence microscopy reconstruction. 
VOLT combines a transport-based formulation that maps degraded measurements to clean volumes via stochastic interpolants with a 3D-native anisotropic network that separates lateral and axial processing. 
This design operates directly in voxel space and achieves improved scalability to large volumes without relying on slice-wise approximations. 
We develop both stochastic (SDE) and deterministic (ODE) variants within the same framework. 
We validate VOLT on simulated wide-field microscopy datasets. Our results show that VOLT significantly improves reconstruction quality in both lateral and axial directions while providing voxel-wise credibility estimates.
\end{abstract}

\section{Introduction}
\label{sec:intro}

Volumetric wide-field fluorescence microscopy has become a major tool for three-dimensional (3D) imaging of thick tissues, organoids, and live-cell networks~\cite{hsieh2025imaging3d,maharjan2024advanced3d}. 
By illuminating the full lateral field simultaneously and stacking focal planes sequentially along the axial direction, it achieves high throughout with simple hardware and low phototoxicity, when compared to laser-scanning techniques such as confocal~\cite{pawley2006handbook} or two-photon microscopy~\cite{helmchen2005deep}.
Despite these strengths, wide-field microscopy lacks the depth selectivity of a confocal pinhole. 
As a result, every acquired focal plane receives contributions from fluorophores above and below the focal plane, producing a \emph{characteristic blur} that degrades both axial and lateral resolution throughout the volume. 
This out-of-focus contamination becomes increasingly severe in thick or densely labeled specimens, where the ratio of in-focus to out-of-focus signal is low~\cite{mertz2019introduction}. 
Therefore, deconvolving the true fluorescence distribution from the degraded measurement constitutes an \emph{ill-posed} inverse problem.

There has been considerable interest in developing computational methods for addressing this challenge. 
Classical approaches~\cite{gonzalez2018digital, orieux2010bayesian, richardson1972bayesian, lucy1974iterative} leverage explicit point-spread function (PSF) knowledge to yield interpretable solutions, but remain sensitive to noise and struggle to recover high-frequency details. 
Deep learning-based methods---including end-to-end mapping~\cite{weigert2018content, sun2018efficient, li2022deep3d} and neural fields~\cite{liu2022recovery, kang2024coordinate}---exploit the strong representation power of neural networks to regularize the solution, yielding high-quality estimates. 
Yet their deterministic nature provides only single-point estimate with no characterization of reconstruction credibility, which is important for drawing robust conclusions in scientific discovery~\cite{ekmekci2023quantifying, ekmekci2025conformalized}.
Recent advances in generative modeling have since introduced probabilistic reconstruction to microscopy~\cite{pan2023diffuseir,li2024physicsinformed}, enabling a principled way to characterize solution uncertainty by reconstructing a set of plausible images consistent with the measurement.
This family encompasses diffusion models~\cite{song2021scorebased,ho2020denoising}, which learn to reverse a pre-defined Gaussian noising process, and flow matching~\cite{lipman2023flow, liu2022flow}, which learns a 
deterministic transport between two distributions. 
However, directly processing 3D volumes within these frameworks remains a computational challenge. 
Existing methods either rely on 2D slice-wise processing that can be suboptimal for axial information recovery~\cite{zhang2018deep_optical_sectioning, pan2023diffuseir, li2026computationaltirfenablesoptical}, or learn a separate image encoder and decoder to operate in a low-dimensional latent space~\cite{rombach2022high, guo2025psi3d}, which can be hard to train.

\begin{figure*}[t!]
\centering
\includegraphics[width=0.95\linewidth]{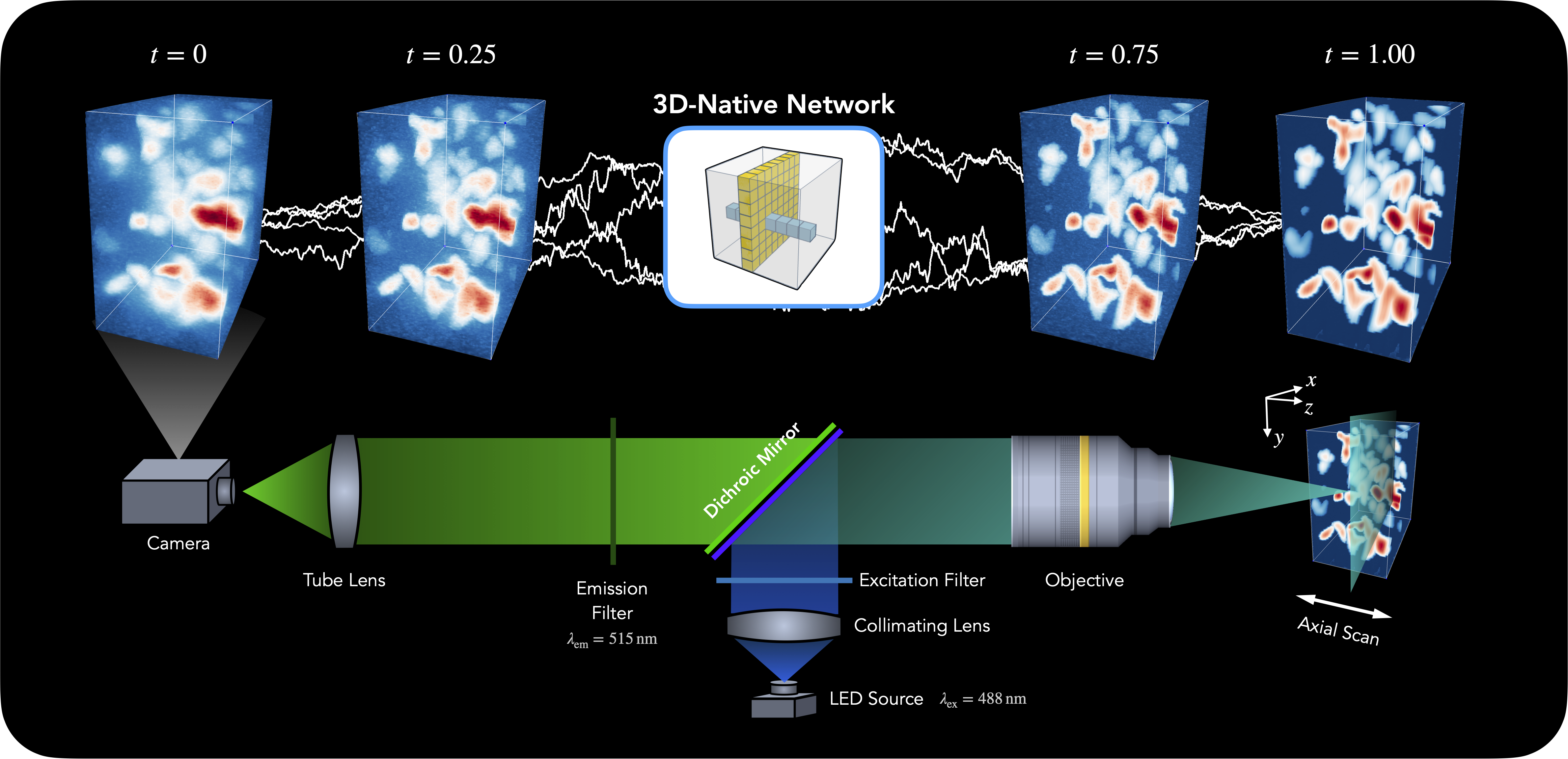}
\caption{Conceptual illustration of the proposed \proposed~framework for wide-field fluorescence microscopy. 
The \emph{bottom portion} illustrates the imaging process, where the camera captures a blurred 3D volume by axially scanning the specimen. 
The \emph{top portion} depicts \proposed, which employs a 3D-native network to probabilistically transport the degraded measurement to a clean volume. 
Unlike existing methods, \proposed~operates directly in pixel space and performs end-to-end volumetric reconstruction while providing uncertainty estimates.
}
\label{fig:scheme}
\end{figure*}

In this work, we propose \emph{volumetric transport (\proposed)}, an \emph{end-to-end, voxel-space} probabilistic method for 3D wide-field fluorescence microscopy reconstruction. \proposed \, makes two core contributions:
\begin{itemize}

    \item \textbf{3D-native anisotropic architecture.} We introduce a volumetric network built from decoupled lateral ($x$-$y$ plane) and axial ($z$-axis) convolutions, inspired by video diffusion architectures~\cite{ho2022video, singer2023makeavideo, wang2023modelscope}.
    This design directly operates in pixel space, naturally accounting for the anisotropic resolution of wide-field microscopy where lateral resolution exceeds axial resolution. 
    The architecture is agnostic to the choice of generative framework---compatible with both diffusion- and flow-based setups---making it a general-purpose backbone for volumetric microscopy reconstruction.

    \item \textbf{Measurement-to-clean probabilistic transport.} Unlike standard diffusion models that initiate transport from a fixed Gaussian distribution, \proposed~leverages \emph{stochastic interpolants}~\cite{albergo2025stochastic,albergo2022building} to directly interpolate between the distributions of degraded measurements and clean volumes. 
    This end-to-end formulation bypasses the need to approximate intractable time-dependent data-fidelity terms for adapting unconditional diffusion models to inverse problems~\cite{chung2023diffusion,kim2025flowdps}.
    The result is an ensemble of high-quality reconstructions, whose voxel-wise variation provides a principled characterization of solution uncertainty.
    
\end{itemize}

We validate our method on simulated wide-field microscopy data generated using a physically accurate 3D PSF model. 
Our results show that \proposed~outperforms standard 3D architectures by offering substantially improved scalability. 
In addition, our qualitative and quantitative comparisons against a diverse set of baselines show that \proposed~achieves higher reconstruction quality in both lateral and axial directions and provides voxel-wise credibility estimates.

\section{Background}
\label{sec:background}

In this section, we introduce the forward model for wide-field fluorescence microscopy and review reconstruction methods ranging from classical model-based methods to deep generative models. 

\subsection{Inverse Problem in Wide-Field Fluorescence Microscopy}

Wide-field fluorescence microscopy illuminates the entire specimen at once, collecting emitted light across the full field of view through the objective lens; Fig.~\ref{fig:scheme} provides a schematic illustration. 
The acquisition process can be described by the 3D emission point spread function (PSF) $h(x,y,z)$, which characterizes how the optical system blurs a single point emitter\cite{NaJi2020CNN_Aberration}
\begin{equation}
\begin{split}
h(x,y,z)
&= \left|
\mathscr{F}
\left[
P(k_x, k_y)\;
\exp\left(i \frac{2\pi}{\lambda_{\textrm{em}}}\,\phi(k_x,k_y)\right)\;
\right.\right. \\
&\qquad\left.\left.\times\exp\left(-2\pi i z \sqrt{\frac{n_0^2}{\lambda_{\textrm{em}}^2} - k_x^2 - k_y^2}\right)
\right]
\right|^2,
\label{eq:psf}
\end{split}
\end{equation}
where $P(k_x, k_y)$ is the complex pupil function, $\phi(k_x,k_y) = \sum_m a_m Z_m(k_x,k_y)$ is the cumulative optical aberration expressed as a weighted sum of Zernike modes, $n_0$ is the immersion medium refractive index, $\lambda_{\textrm{em}}$ is the fluorescence emission wavelength, and $\mathscr{F}$ denotes the 2D Fourier transform over pupil coordinates. 

Discretizing $h(x,y,z)$ onto the imaging grid yields the forward model
\begin{equation} \label{eq:inv}
    \ybf = \hbf * \xbf + \bm{\xi},
\end{equation}
where $\xbf \in \R^n$ is the underlying fluorophore distribution, $\ybf \in \mathbb{R}^n$ is the observed blurry volume, $*$ denotes 3D convolution, and $\bm{\xi} \in \R^n$ models mixed Poisson shot noise and Gaussian detector read-out noise. 
The inverse problem of recovering $\xbf$ from $\ybf$ is ill-posed due to the low-pass filtering effect of $\hbf$ and the presence of noise.
A defining characteristic of wide-field microscopy is its anisotropic resolution, where lateral resolution is significantly higher than axial resolution due to the elongated point spread function along the optical axis. 
This anisotropy leads to non-uniform degradation across dimensions, with stronger blur in the axial direction. 
This observation motivates the design of \proposed, which explicitly separates lateral and axial processing for efficiency while still preserving reconstruction quality.

\subsection{Model-Based Deconvolution}
Classical linear filters such as Wiener filtering~\cite{gonzalez2018digital} and constrained least-squares (CLS)~\cite{orieux2010bayesian} invert the optical transfer function in the frequency domain given a known PSF; however, they amplify noise and produce ringing artifacts.
The Richardson--Lucy (RL) algorithm~\cite{richardson1972bayesian, lucy1974iterative} iteratively maximizes the Poisson likelihood and has become a standard tool in microscopy software~\cite{biggs20103d, mcnally1999threedimensional}, yet requires careful early stopping and explicit PSF knowledge.
Total-variation~\cite{dey2006richardsonlucy} and sparsity-based regularizers improve robustness to noise, but all such approaches remain tightly coupled to explicit forward-model assumptions and struggle to recover fine structural detail under severe axial blur.

\begin{figure*}[t!]
    \centering
    \includegraphics[width=0.95\linewidth]{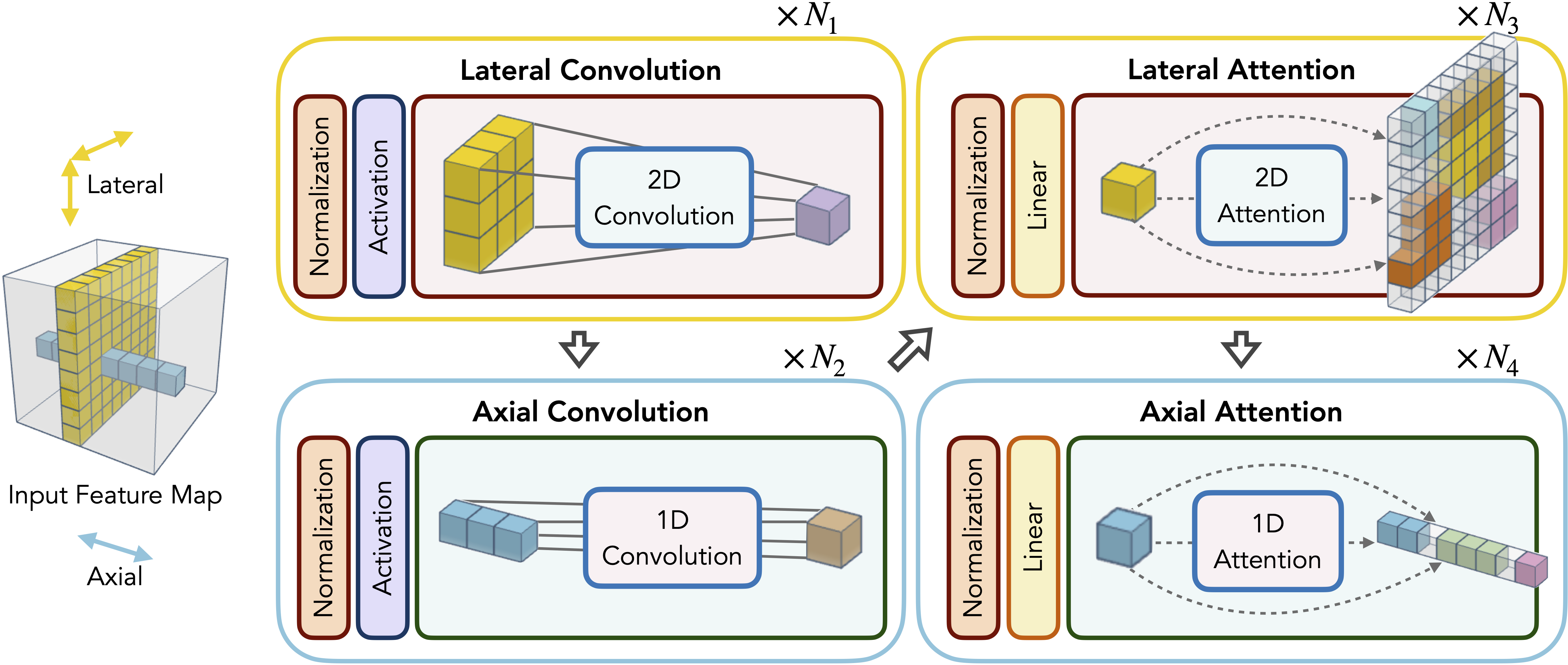}
    \caption{
    Visual illustration of the proposed lateral--axial factorization. Unlike standard 3D convolutions, our architecture applies lateral and axial convolutions separately, followed by attention modules along their respective dimensions. This factorization improves scalability by enabling deeper networks under the same GPU memory budget; see comparison in Fig.~\ref{fig:memory}.
    }
    \label{fig:arch}
\end{figure*}

\subsection{Deterministic Deep Learning}
Deep learning has substantially advanced inverse problem solving by exploiting the strong representation power of neural networks for regularizing solutions~\cite{jin2017deep, sun2018efficient, weigert2018content, gilton2021deep}.
In the context of computational microscopy, a line of research~\cite{weigert2018content, sun2018efficient} trains encoder--decoder convolutional neural networks to map degraded volumes to clean counterparts in an end-to-end manner, given paired training data. 
Subsequent 3D architectures~\cite{li2022deep3d,wolny2020accurate} further improve quality by incorporating volumetric context. 
Neural fields (NFs)~\cite{mildenhall2021nerf, liu2022recovery,kang2024coordinate} offer an alternative paradigm, representing the underlying volume as a coordinate-based network trained to fit the observed measurement through a differentiable forward model. 
Unlike end-to-end methods, NFs require no training dataset, but have to be re-optimized for every new test volume. 
While both families of methods yield high-quality reconstructions in their respective favorable settings, they are inherently deterministic and provide no mechanism for characterizing solution uncertainty.

\subsection{Transport-based Generative Models and Reconstruction}
Recent advances in generative diffusion models~\cite{ho2020denoising} have spurred rapid development in transport-based generative modeling. 
Existing approaches can be broadly categorized by how they construct transport between distributions, either through \emph{stochastic differential equations (SDEs)} or \emph{ordinary differential equations (ODEs)}, as well as by the choice of endpoint distributions. 
SDE-based methods, exemplified by diffusion models~\cite{song2021scorebased}, stochastically connect a simple reference distribution, such as Gaussian noise, to the data distribution. 
In contrast, ODE-based approaches, such as flow matching~\cite{lipman2023flow, liu2022flow}, learn deterministic transport maps via continuous-time velocity fields, but yield a single-point reconstruction and therefore do not capture trajectory diversity. 
Both classes have been adapted to inverse problems, either by incorporating measurement-consistency constraints during sampling~\cite{chung2023diffusion, wu2024principled} or by matching the generation process to the reconstruction dynamics~\cite{delbracio2023inversion,kawar2022denoising}.
More recently, the framework of stochastic interpolant~\cite{albergo2022building, albergo2025stochastic} unifies these perspectives by defining continuous-time bridges between arbitrary endpoint distributions with tunable stochasticity, recovering both stochastic and deterministic transport as special cases.

Applying these frameworks to 3D microscopic imaging remains challenging, as the underlying networks must process high-dimensional volumetric data, making full 3D training computationally expensive. 
\proposed~combines efficient lateral-axial 3D processing with stochastic interpolants to learn an end-to-end transport from degraded measurements to clean volumes, enabling probabilistic reconstruction with voxel-wise uncertainty.

\section{Proposed Method: \proposed}
\label{sec:\proposed}

\proposed~introduces a 3D-native network architecture for processing microscopic volumes with anisotropic resolution, enabling direct transport between the distributions of degraded measurements and that of clean volumes. 
We begin with the network architecture and then explain how the transport can be achieved based on the stochastic interpolant framework.

\subsection{Lateral-Axial Factorization for Efficient 3D-Native Processing} 
\label{sect:network}

Standard 3D convolutions handle volumetric inputs by processing all dimensions jointly. 
However, this makes the overall network scale poorly when combined with attention mechanisms that are essential yet computationally expensive.
Inspired by the spatio-temporal decomposition in video generation~\cite{ho2022video, singer2023makeavideo, wang2023modelscope}, \proposed~introduces lateral-axial blocks, each consisting of four sequentially applied modules:
\begin{enumerate}
    \item \textbf{Lateral convolution.} Standard 2D convolutions applied independently to each $z$-slice, extracting in-plane spatial features such as edges, textures, and structures.
    \item \textbf{Axial convolution.} 1D convolutions applied along the $z$-axis at each lateral position, modeling local inter-slice dependencies and short-range axial correlations.
    \item \textbf{Lateral attention.} Multi-head self-attention over spatial tokens within each $z$-slice, capturing long-range lateral structure.
    \item \textbf{Axial attention.} Multi-head self-attention across $z$-slices at each spatial position, capturing long-range depth-dependent correlations.
\end{enumerate}

Fig.~\ref{fig:arch} illustrates the proposed lateral-axial blocks. 
This factorization reduces computational cost relative to full 3D convolutions and attention (see Sec.~\ref{sect:memory} \& Fig.~\ref{fig:memory} for details), by allowing the network to learn distinct representations for lateral and axial dimensions. 
While slice-wise approaches~\cite{pan2023diffuseir, guo2025psi3d} adopt a related philosophy, they typically apply a generative model only to individual 2D slices and handle the axial dimension via hand-crafted regularizers such as total variation. 
In contrast, our factorization couples lateral and axial processing directly within the architecture, enabling a native 3D volumetric representation that captures inter-slice correlations.

\subsection{Measurement-to-Clean Probabilistic Transport}

Rather than seeking a single point estimate, \proposed~formulates the 3D deconvolution problem~\eqref{eq:inv} as a transport from the distribution $\rho_0$ of degraded measurements to the distribution $\rho_1$ of clean fluorescence volumes. 
We instantiate this transport using the stochastic interpolant framework~\cite{albergo2025stochastic}. 
For notational clarity, we denote $\mathbf{x}_0 = \mathbf{y} \sim \rho_0$ and $\mathbf{x}_1 = \mathbf{x} \sim \rho_1$ throughout this section.

We consider the linear interpolant between $\xbf_0$ and $\xbf_1$
\begin{equation}
    \label{eq:interpolant}
    \mathbf{x}_t = \alpha_t\,\mathbf{x}_0 + \beta_t\,\mathbf{x}_1 + \gamma_t\,\mathbf{z}, \quad t \in [0,1],
\end{equation}
where $\mathbf{z} \sim \mathcal{N}(\mathbf{0}, \mathbf{I})$ is an independent latent variable that regularizes the intermediate distribution. 
The schedule functions $\alpha_t$, $\beta_t$, and $\gamma_t$ are differentiable and satisfy the boundary conditions $\alpha_0 {=} \beta_1 {=} 1$ and $\alpha_1 {=} \beta_0 {=} \gamma_0 {=} \gamma_1 {=} 0$, ensuring that $\mathbf{x}_t$ recovers the degraded volume at $t{=}0$ and the clean volume at $t{=}1$. 
We adopt $\alpha_t = 1 - t$, $\beta_t = t$, and $\gamma_t = 0.1\sin^2(\pi t)$, which maintains stable derivatives near the endpoints.

\proposed~generates clean volumes from degraded measurements by integrating the SDE associated with~\eqref{eq:interpolant}
\begin{equation}
    d\mathbf{x}_t = \Big[b(\mathbf{x}_t,t) + \epsilon(t)\,s(\mathbf{x}_t,t)\Big]\,dt + \sqrt{2\epsilon(t)}\,d\mathbf{W}_t,
    \label{eq:sde}
\end{equation}
from $t{=}0$ to $t{=}1$, initialized at $\mathbf{x}_0 = \mathbf{y}$. 
The marginal velocity field $b(\mathbf{x}_t, t)$ and score function $s(\mathbf{x}_t, t)$ characterize the intermediate density $\rho_t$ associated with $\mathbf{x}_t$
\begin{align}
b(\mathbf{x}_t, t) &= \dot{\alpha}_t\,\mathbb{E}[\mathbf{x}_0 \mid \mathbf{x}_t] + \dot{\beta}_t\,\mathbb{E}[\mathbf{x}_1 \mid \mathbf{x}_t] + \dot{\gamma}_t\,\mathbb{E}[\mathbf{z} \mid \mathbf{x}_t], \label{eq:volecity} \\
s(\mathbf{x}_t, t) &= \nabla_{\mathbf{x}_t} \log p_t(\mathbf{x}_t) = -\frac{1}{\gamma_t}\,\mathbb{E}[\mathbf{z} \mid \mathbf{x}_t], \label{eq:score}
\end{align}
where~\eqref{eq:score} follows from the Gaussian conditional structure of~\eqref{eq:interpolant}, and $\dot{\alpha}_t$, $\dot{\beta}_t$, $\dot{\gamma}_t$ denote time derivatives of the schedules. 
In~\eqref{eq:sde}, the diffusion coefficient $\epsilon(t) \geq 0$ controls the level of stochasticity. Setting $\epsilon(t) = 0$ reduces~\eqref{eq:sde} to the ODE flow $d\mathbf{x}_t = b(\xbf_t, t)\,dt$; 
further letting $\gamma_t = 0$ recovers deterministic flow matching~\cite{lipman2023flow}. 
We emphasize that $\epsilon(t)$ and $\gamma_t$ play distinct roles. 
The former governs the stochasticity of the sampling dynamics, while the latter determines the structure of the interpolant distribution $\rho_t$. 
For numerical convenience, we set $\epsilon(t) = 0.1 \sin^2(\pi t)$, using the same schedule as $\gamma_t$.
In Sec.~\ref{sec:results}, we develop both \proposed~(SDE) and \proposed~(ODE) and validate their effectiveness, where the SDE variant draws diverse samples given the same measurement. 

Given that both velocity and score are analytically intractable, we approximate them using a velocity network $b \approx b_\theta$ and a score network $s \approx s_\phi$, each implemented with the proposed lateral-axial architecture for efficient 3D volume processing. 
These networks are trained by minimizing the following objectives~\cite{albergo2025stochastic}
\begin{align}
    \mathcal{L}_\theta &= \E_{t,\xbf_0,\mathbf{x}_1,\zbf}\|b_\theta(\xbf_t,t) - \big(\dot\alpha_t\mathbf{x}_0 + \dot\beta_t\mathbf{x}_1 + \dot\gamma_t\zbf\big)\|^2, \label{eq:loss_b}\\
    \mathcal{L}_\phi &= \E\|s_\phi(\mathbf{x}_t,t) - (-\mathbf{z}/\gamma_t)\|^2, \label{eq:loss_eta}
\end{align}
with $t \sim \mathsf{Uniform}(0,1)$. 
To avoid numerical instability near the boundary where $\gamma_t \to 0$, we parameterize the score as $s_\phi = \eta_\phi / \gamma_t$, where $\eta_\phi$ is a network predicting $-\zbf$.

\begin{table*}[t!]
\renewcommand{\arraystretch}{1.3}
\caption{Numerical Results for \proposed, standard SI, and baselines averaged over $80$ test volumes. For \proposed~(SDE) and SI (SDE), we average the metrics over $5$ independent samples for each test volume and report the mean and the standard deviation. The best results are in \colorbox{hlcolor}{\textbf{bold}}, and the second best are \underline{underlined}.}
\centering
\label{table}
\footnotesize
\setlength{\tabcolsep}{10pt}
\begin{tabular}{l cccc}
\toprule
Methods& 
PSNR$\uparrow$&
SSIM$\uparrow$& 
MS-SSIM$\uparrow$&
LPIPS$\downarrow$\\
\midrule
3D Flow Matching& 28.541& 0.947& 0.967& 0.058\\
CoCoA& 24.749& 0.907& 0.789& 0.168\\
3D U-Net& 27.560& 0.929& 0.962& 0.076\\
2D U-Net& 21.966& 0.908& 0.824& 0.059\\
RL& 19.648& 0.883& 0.642& 0.255\\
CLS& 14.620& 0.041& 0.101& 0.429\\

\midrule
SI (SDE) & \underline{29.626} $\pm$0.065 & \underline{0.958} $\pm$0.001 & \underline{0.972} $\pm$0.001 & 0.053 $\pm$0.001 \\
SI (ODE) & 27.370 & 0.917 & 0.956 & 0.077 \\

\midrule
\proposed \, (SDE) & \colorbox{hlcolor}{\textbf{29.855}} $\pm$0.073 & 0.953 $\pm$0.001 & 0.965 $\pm$0.001 & \colorbox{hlcolor}{\textbf{0.037}} $\pm$0.001 \\
\proposed \, (ODE) & 29.218 & \colorbox{hlcolor}{\textbf{0.959}} & \colorbox{hlcolor}{\textbf{0.972}} &\underline{0.041}\\

\bottomrule
\end{tabular}
\label{metric_table}
\end{table*}

\subsection{Low-Dimensional Reparameterization}

Recent work~\cite{kaimingHe_predictx1_2025} suggests that transport-based methods can achieve improved predictions by directly estimating the clean image rather than the velocity or score, as the former mainly lie on a lower-dimensional manifold while the latter often contain high-frequency components such as noise.
Inspired by this observation, we reparameterize the velocity network to predict the clean volume $\hat{\xbf}_1 = b_\theta(\mathbf{x}_t, t)$, using a more direct objective on the data manifold
\begin{equation}
    \mathcal{L}_\theta^+ = \E \| \hat{\mathbf{x}}_1 - \mathbf{x}_1 \|^2 + \lambda \cdot (1-\mathrm{SSIM}(\hat{\mathbf{x}}_1 - \mathbf{x}_1))
    \label{eq:loss_b_x1}
\end{equation}
where $\mathrm{SSIM}$ is the \emph{structural similarity index measure}~\cite{wang2004image} with regularization strength $\lambda$; we choose $\lambda = 1$ in our experiments. 
For sampling, we can analytically recover the velocity field by substituting $\mathbf{z} = (\mathbf{x}_t - \alpha_t \mathbf{x}_0 - \beta_t \hat{\mathbf{x}}_1)/\gamma_t$ into the conditional velocity
\begin{equation}
    b_\theta(\mathbf{x}_t, t) = \dot{\alpha}_t\, \mathbf{x}_0 + \dot{\beta}_t\, \hat{\mathbf{x}}_1 + \frac{\dot{\gamma}_t}{\gamma_t}\big(\mathbf{x}_t - \alpha_t\, \mathbf{x}_0 - \beta_t\, \hat{\mathbf{x}}_1\big)\,.
    \label{eq:x1_reparam}
\end{equation}
We show ablation studies in Table~\ref{table}, where SI variants in the middle section denote classical stochastic interpolant implementations. Compared to SI (ODE) trained with standard velocity field regression, \proposed~(ODE) improves PSNR by approximately \SI{1.8}{\decibel} and reduces LPIPS by 47\%. Low-dimensional reparameterization also improves the SDE variant, where sample means from \proposed~(SDE) achieve \SI{0.2}{\decibel} higher PSNR than SI (SDE). This demonstrates that both ODE and SDE benefit from direct regression on the clean volume manifold compared to velocity-space regression.

\begin{figure*}[t!]
    \centering
    \includegraphics[width=1\linewidth]{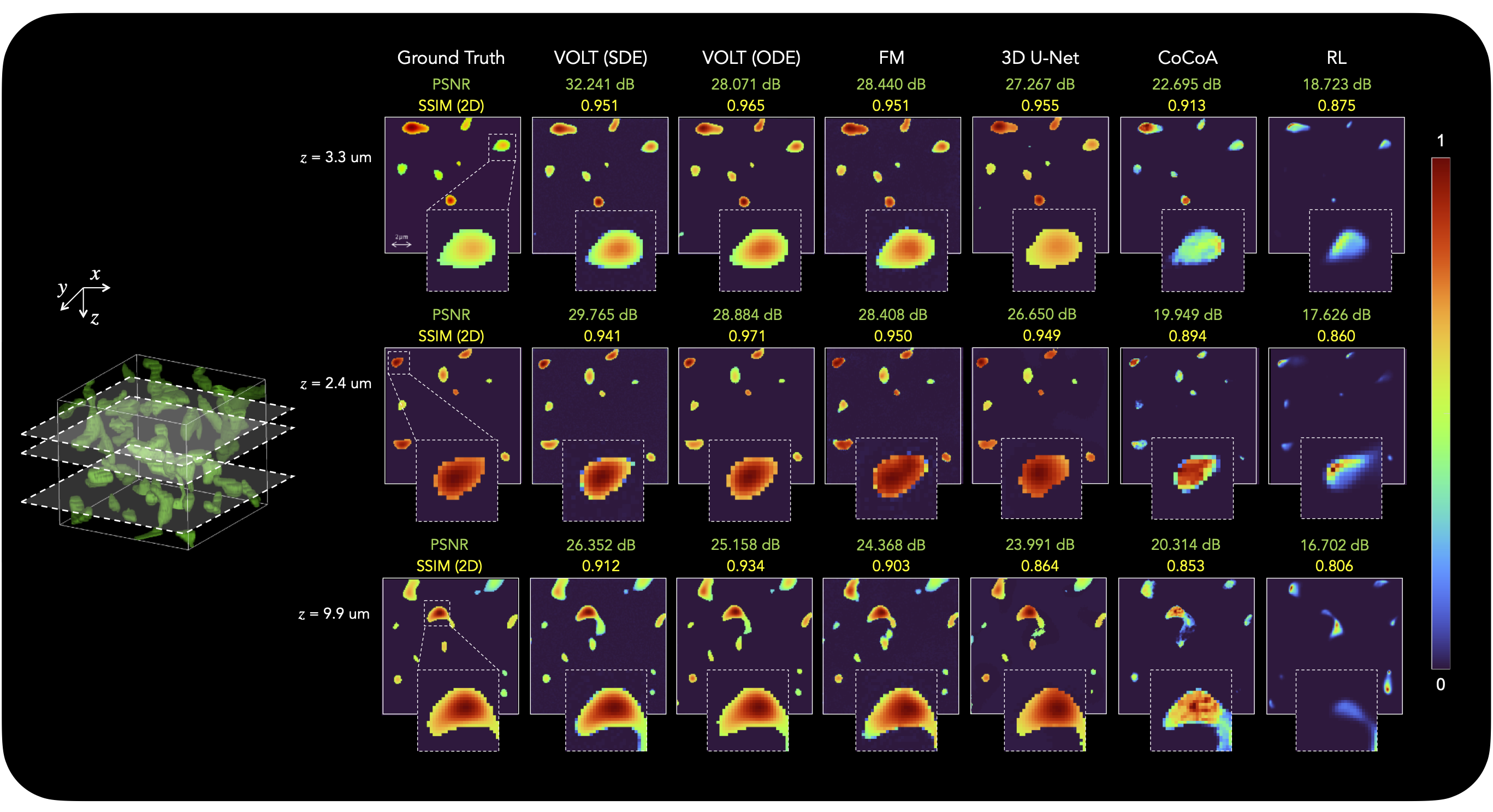}
    \caption{
    Visual comparison of lateral ($x$-$y$) reconstructions for an example test volume. 
    Three slices at $z = 3.3 \, \mu m$, $z =2.4 \, \mu m$, and $z = 9.9 \, \mu m$ are shown in the three rows, along with magnified insets and per-slice PSNR/SSIM values. 
    Note how \proposed~(SDE) and \proposed~(ODE) best preserve the structural morphology and intensity distribution of the ground truth.}
    \label{fig:wide_comparison}
\end{figure*}

\section{Experiment \& Results}
\label{sec:results}

In this section, we validate \proposed~on simulated wide-field fluorescence microscopy data. We first describe the problem setup and implementation details, then present numerical comparisons, followed by lateral and axial reconstruction comparisons. We also include an analysis of attention memory scalability and reconstruction credibility.

\subsection{Problem Setup \& Implementation Details}

We follow classical 3D deconvolution microscopy protocols as outlined in~\cite{biggs20103d, NaJi2020CNN_Aberration, kang2024coordinate}. Consider a diffraction-limited wide-field epi-fluorescence microscope with an excitation wavelength of $\lambda_{\textrm{ex}} {=}$\SI{488}{\nano\meter} and a water-immersed objective lens with numerical aperture $\mathrm{NA}{=}1.1$. The refractive index of the immersion medium (water) is $n_0 {=} 1.33$. We assume that the fluorescence wavelength of the sample is $\lambda_{\textrm{em}} {=}$\SI{515}{\nano\meter}, which is the typical peak emission wavelength of \emph{fluorescein isothiocyanate (FITC)} widely used in fluorescence microscopy. We set Zernike coefficients $a_m = 0$ for all non-piston modes so that $\phi(k_x, k_y) = \mathrm{constant}$ in~\eqref{eq:psf}. This aberration-free approximation is valid for small to moderate fields of views and depths of views.

We simulate clean fluorophore distributions $\xbf$ by generating random continuous gradient fields with simplex noise \cite{simplexNoise2002}, before randomly thresholding them to obtain irregular cell structures with various sizes, shapes, and densities. Each segmented cell is then scaled by a random base intensity. We generate $800$ samples of $136\times 136\times 40$ and divide them into $640$ training, $80$ validation, and $80$ testing volumes. We choose the sampling grid $(dx, dy, dz)$ as $(0.1, 0.1, 0.3)$ \SI{}{\micro\meter}, yielding a sample volume $\xbf$ of $13.6 \times 13.6\times 12$ \SI{}{\micro\meter\cubed}, where the aberration-free PSF assumption is sufficient.

For the \proposed~framework, we construct two anisotropic 3D U-Nets with lateral-axial blocks detailed in Sec~\ref{sect:network} to model the velocity field and the score function.  
Each lateral-axial block consists of $2$ lateral convolutions, $4$ axial convolutions, $2$ lateral attention, and $2$ axial attention layers. 
All attention modules are implemented with FlashAttention~\cite{dao2022flashattention}.
We use U-Net architectures with $3$ lateral downsamplings to bring $x$-$y$ dimensions from $136$ to $17$, while keeping the axial dimension fixed at $40$. 
We use low-dimensional reparameterization in~\eqref{eq:x1_reparam} for the velocity field, and train the two networks with time-indexed paired data $(\mathbf{x}_t, \mathbf{x}_0, t)$ where the time is injected via sinusoidal encodings followed by a multilayer perceptron~\cite{dhariwal2021diffusion}. For a given degraded input volume $\mathbf{y}$, we initialize $\mathbf{x_0} = \mathbf{y}$ and integrate the forward SDE~\eqref{eq:sde} from $t = 0$ to $t = 1$ using a second-order Heun sampler with $100$ steps to obtain the clean fluorescence distribution $\mathbf{x} = \mathbf{x}_1$. We train the 3D U-Net networks using AdamW optimizer with an initial learning rate of $5 \times 10^{-5}$, followed by a reduction factor of $0.2$ once the loss plateaus. We train and evaluate \proposed~on a single NVIDIA H200 GPU. 

\subsection{Numerical Comparison}

In this section, we compare our method to several classical and learning-based approaches, including \textit{(a)} linear filtering with CLS and Laplacian regularization~\cite{orieux2010bayesian}, \textit{(b)} standard RL deconvolution\cite{richardson1972bayesian,lucy1974iterative}, where we select an optimal number of iterations, \textit{(c)} end-to-end 2D U-Net from~\cite{wolny2020accurate}, applied slice-wise to our 3D sample, \textit{(d)} end-to-end 3D U-Net~\cite{wolny2020accurate}, \textit{(e)} NF-based methods such as CoCoA~\cite{kang2024coordinate}, and \textit{(f)} flow matching~\cite{lipman2023flow}, where we adopt the basic coupling between distributions by setting $\gamma_t = 0$ in~\eqref{eq:interpolant} and learn a velocity field with the same anisotropic U-Net used in our method. We show the overall numerical evaluation, as well as lateral and axial reconstruction fidelity, respectively.

Table~\ref{table} shows numerical evaluation averaged over $80$ test volumes using four metrics: \emph{peak signal-to-noise ratio (PSNR)}, \emph{SSIM}, \emph{multi-scale SSIM (MS-SSIM)}~\cite{wang2003multiscale}, and \emph{learned perceptual image patch similarity (LPIPS)}~\cite{zhang2018unreasonable}, where LPIPS is computed slice-wise and averaged. 
For \proposed~(SDE), we average each metric over $5$ independent sampling runs per test volume to account for stochasticity and report the standard deviation. 
Both \proposed~variants substantially outperform all remaining baselines. \proposed~(SDE) achieves the best PSNR performance, achieving \SI{1.3}{\decibel} improvement in PSNR over 3D flow matching, and \SI{2.3}{\decibel} improvement in PSNR over 3D U-Net. The small standard deviations in the metrics demonstrate that \proposed~(SDE) consistently outputs high-quality samples. \proposed~(ODE) closely follows \proposed~(SDE) with nearly identical results and slightly higher SSIM and MS-SSIM. 

In practice, the two \proposed~variants offer complementary trade-offs. 
\proposed~(SDE) provides the highest reconstruction fidelity by integrating~\eqref{eq:sde} over 100 Heun steps, while providing spatially-resolved credibility maps via multiple independent samples. On the other hand, \proposed~(ODE) achieves nearly identical metrics in PSNR and slightly better perceptual quality using only 20 ODE steps, which is well-suited for applications that require efficient and deterministic reconstruction.

For the other baselines, the NF-based CoCoA yields relatively high PSNR and SSIM, but it does not utilize rich information from the training data. 
3D~U-Net achieves moderate SSIM but substantially lower PSNR (\SI{27.6}{\decibel}) and poor MS-SSIM. The slice-wise 2D U-Net performs worse due to the absence of inter-slice consistency. 
Model-based methods such as RL deconvolution and CLS filtering perform worst overall due to the lack of learned structural information and their sensitivity to ill-posedness. 

\subsection{Lateral Reconstruction Comparison}
Fig.~\ref{fig:wide_comparison} shows an example test volume on the left and lateral ($x$-$y$) slices extracted at $z = 3.3 \, \mu m$, $z =2.4 \, \mu m$, and $z = 9.9 \, \mu m$ in each row. 
Each column represents a different reconstruction method, along with magnified insets and per-slice PSNR and SSIM values. \proposed~(SDE) and \proposed~(ODE) best recover the structural details in the ground truth~(GT), preserving both the morphology and intensity distribution of the fluorescent structures. By contrast, other learning-based methods such as 3D U-Net produce flattened structures with a visible loss of fine textural and morphological detail. This result is consistent with the tendency of convolutional neural networks trained with per-voxel regression losses to predict over-smoothed reconstructions~\cite{bishop2006pattern}.

We note an important difference between the two \proposed~variants. 
\proposed~(ODE) reconstructions exhibit jagged structural boundaries, whereas \proposed~(SDE) produces smoother and more coherent shapes. \proposed~(ODE) only learns a velocity field $b_\theta$; approximation errors in $b_\theta$ therefore accumulate along the integration path without correction. \proposed~(SDE), however, employs the full SDE~\eqref{eq:sde} where the score $s = \nabla_{\mathbf{x}_t}\log \rho_t$ provides a corrective drift that steers samples toward high-density regions of the intermediate distribution $\rho_t$. \proposed~(SDE) therefore smooths out trajectory-level artifacts and produces more natural structural boundaries. 

For model-based approaches, Richardson-Lucy (RL) deconvolution fails to recover several structures visible in the ground truth. The RL reconstruction is noticeably shrunken and fragmented, which is consistent with RL being an iterative maximum-likelihood solver. Without the structural priors in \proposed, RL tends to concentrate intensity into isolated peaks, while amplifying high-frequency noise and suppressing weak, low-contrast signals. CoCoA~\cite{kang2024coordinate} partially mitigates these issues with implicit regularization in its NF. However, CoCoA reconstruction quality remains limited compared to supervised methods, as it only relies on a single measurement without access to statistical priors.

\begin{figure}
    \centering
    \includegraphics[width=0.7\linewidth]{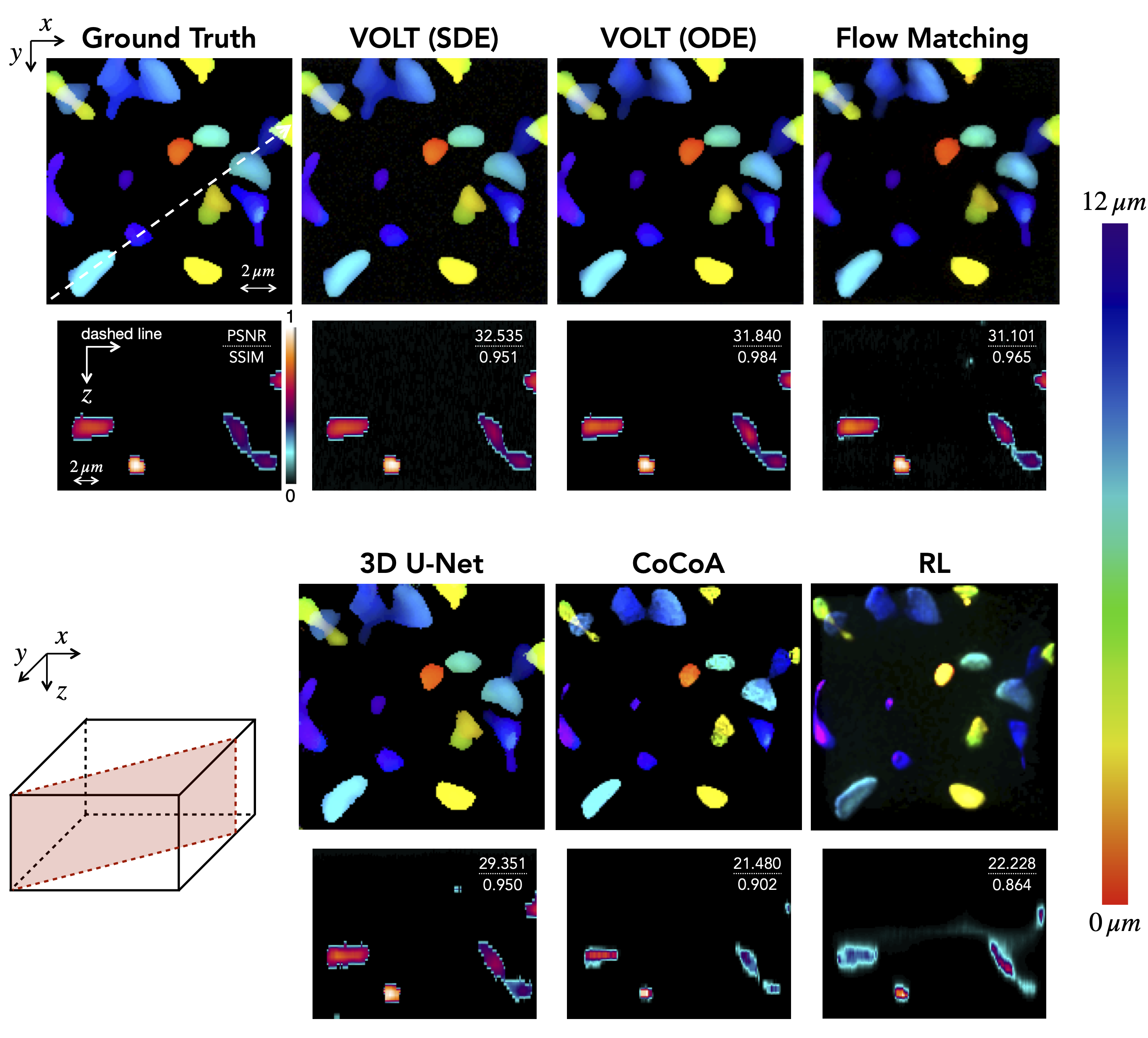}
    \caption{
    Visual comparison of axial reconstructions for a test volume. Each maximum intensity projection (MIP) is color-coded with axial depth, and plotted along with a cross-sectional view extracted along the dashed line shown in the ground truth. The cross-sectional view is further annotated with the corresponding PSNR/SSIM values. We also show a schematic on the bottom left to illustrate the cross-sectional plane.     
    Note how \proposed~accurately recovers axial profiles over the baselines.}
    \label{fig:mip}
\end{figure}

\subsection{Axial Reconstruction Comparison}
For each reconstruction method, we show depth-color-coded \emph{maximum intensity projection (MIP)} maps in Fig.~\ref{fig:mip}, where the colorbar on the right encodes depth in the axial direction. For each MIP map, we also show a corresponding cross-sectional view extracted along the dashed lines in the GT map, displayed using the GT cross-section colormap. We also show a schematic on the lower left to illustrate the cross-sectional plane relative to the volume. In the MIP maps, all supervised learning-based methods yield a lateral reconstruction broadly consistent with the ground truth, whereas RL and CoCoA suffer from missing or shrunken structures. However, the cross-sectional views reveal some more discriminating information, where only transport-based methods, including \proposed~and flow matching, correctly recover axial profiles that match the ground truth. All other baselines show severe axial elongation or compression. Notably, 3D-UNet, despite having paired training data, shows poor axial fidelity with visibly distorted, jagged cross-sectional shapes. This suggests that isotropic 3D convolutions and attentions are insufficient to learn the highly ill-conditioned inverse problem in 3D microscopy. With our anisotropic architecture, \proposed~and flow matching effectively recover the missing axial information.

\subsection{Memory Scalability}
\label{sect:memory}

\begin{figure}
    \centering
    \includegraphics[width=0.7\linewidth]{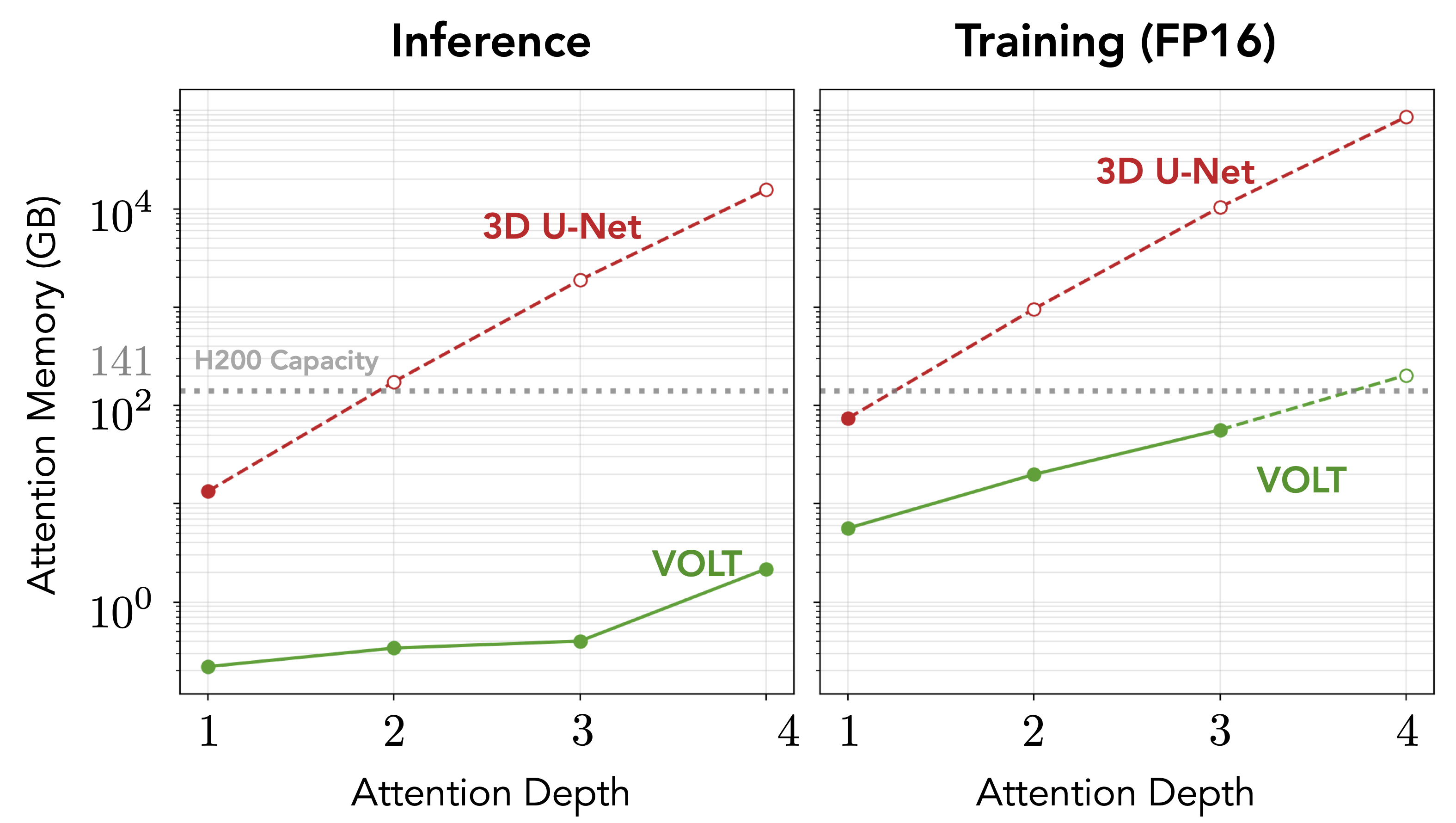}
    \caption{Attention memory overhead (log scale) as a function of attention depth for three architectures. Solid dots are measured on an H200 GPU; hollow dots are extrapolated. The gray dashed line marks the \SI{141}{\giga\byte} H200 capacity. Compared to volumetric attention, \proposed~has a much better memory scalability.}
    \label{fig:memory}
\end{figure}

In this section, we compare the memory scalability of VOLT against the standard 3D U-Net with volumetric attention. 
Fig.~\ref{fig:memory} shows the attention memory overhead of the two architectures as a function of attention depth, where depth $d$ indicates that attention modules are placed in the deepest $d$ feature map levels of the network. 
The filled dots are attention memory overhead determined experimentally, while the hollow dots indicate extrapolated values. Note that shallower levels operate on larger feature maps, so memory grows rapidly as $d$ increases. 
The practical impact is substantial: at attention depth~2, naive volumetric attention (U-Net 3D) already exceeds the \SI{141}{\giga\byte} capacity of an H200 GPU during both inference and FP16 training. In contrast, \proposed~remains well below the memory ceiling for all depths during inference, and only slightly exceeds the limit at depth~4 during training. This scalability allows \proposed~to incorporate attention at multiple resolution levels with minimal memory bottlenecks, which is critical for accurate volumetric reconstructions.

\subsection{Reconstruction Credibility}

\begin{figure}
    \centering
    \includegraphics[width=0.7\linewidth]{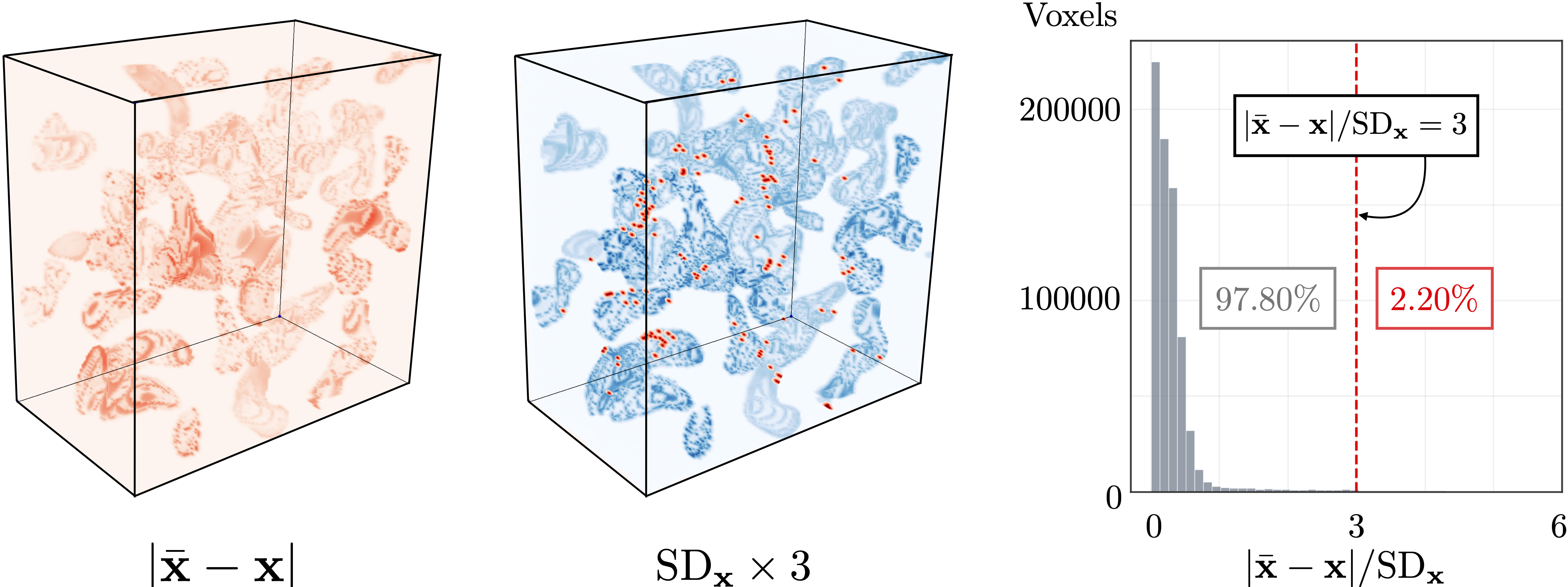}
    \caption{Visualization of voxel-wise statistics enabled by \proposed. 
    We draw 10 samples for the test volume and compute the absolute error $|\bar{\xbf} - \xbf|$ (\emph{left}) and the sample standard deviation (scaled by $\times 3$, \emph{middle}). 
    Red dots indicate voxels where $|\bar{\xbf} - \xbf| / \mathrm{SD}_{\xbf} > 3$.
    Note the scarce distribution of the outliers. 
    A histogram of the per-voxel ratio (\emph{right}) is additionally shown, which summarizes the overall 3-SD coverage.}
    \label{fig:uq}
\end{figure}

A key benefit of \proposed~(SDE) is its ability to characterize reconstruction credibility through repeated sampling. Given a single measurement $\ybf$, we draw $10$ independent samples by integrating~\eqref{eq:sde} with independent Wiener process realizations, yielding a per-voxel sample mean $\bar{\xbf}$ and sample standard deviation $\mathrm{SD}_{\xbf}$. Fig.~\ref{fig:uq} visualizes reconstruction credibility for one test volume. The left panel shows the absolute reconstruction error $|\bar{\xbf} - \xbf|$, while the middle panel displays the sample standard deviation (scaled by 3 for visibility), with red dots marking voxels where $|\bar{\xbf} - \xbf|/\mathrm{SD}_{\xbf} > 3$. The two volumes are visually highly correlated: regions with large reconstruction error consistently coincide with regions of high sample standard deviation, suggesting that sampling standard deviation is a good indicator for reconstruction difficulty. The right panel shows a histogram of the per-voxel ratio $|\bar{\xbf} - \xbf|/\mathrm{SD}_{\xbf}$, where 97.80\% of voxels fall below the $3 \, \mathrm{SD}$ threshold. To assess consistency across the dataset, we repeat this analysis over 10 test volumes, each sampled 10 times. On average, 95.74\% of voxels satisfy the $3 \, \mathrm{SD}$ criterion, with a mean negative log-likelihood (NLL) of $-10.61$. These statistics indicate that \proposed~(SDE) yields diverse yet accurate reconstructions whose spread reliably reflects the underlying reconstruction credibility.

\section{Conclusion}

We presented \proposed, a voxel-space probabilistic transport framework for volumetric wide-field fluorescence microscopy reconstruction. \proposed~decouples lateral and axial dimensions to impose a microscopy-informed inductive bias, while achieving linear memory scaling and enabling volumetric attention implementations under practical GPU budgets. It constructs a direct measurement-to-clean transport via stochastic interpolants and low-dimensional reparameterization. 
Experimental results show that \proposed~(SDE) outperforms a diverse set of baselines while enabling voxel-wise characterization of reconstruction uncertainty. \proposed~(ODE) achieves similar reconstruction quality with fewer sampler steps. 
Future work includes extending VOLT to other microscopy modalities and evaluating its performance on experimentally collected datasets.

 \section*{Acknowledgment}
 This work is partially supported by grant NSF CCF-2542022 and by the Department of Energy Computational Science Graduate Fellowship under Award Number DE-SC0026073.

\end{document}